# Effect of some plant extracts on hardwood cuttings of Bottlebrush (*Callistemon viminalis*)

[1]Hemn Abdalla Mustafa, [1]Tariq Abubakr Ahmad, [1]Aram Akram Mohammed, [1]Zainab Sabah Lazim, [1]Chopi Omer Ibrahim, [1]Roshna Faeq Kak bra, [1]Shvan Ramzi Salih

[1]Horticulture Department, College of Agricultural Engineering Sciences, University of Sulaimani, Kurdistan Region-Iraq

Correspondence: aram.hamarashed@univsul.edu.iq

**Abstract**

The study was conducted at the Collage of Agricultural Engineering Sciences, University of Sulaimani, Kurdistan Region-Iraq so as to investigate response hardwood cuttings of *Callistemon viminalis* to some plant extracts. The hardwood cuttings were taken on 11 March 2021 and soaked separately in 3 and 6 g/L aqueous extracts of moringa leaf, licorice root, willow shoot, fenugreek seed and cinnamon bark for 1 hour. They were compared to the cuttings dipped in 3000 ppm IBA for 10s and control cuttings which were soaked in distilled water for 1 hour. The experiment laid out in CRD with three replications in a greenhouse, and each replication included six cuttings which planted in a mixture of sand and rice husk medium. After 4 months, the measurements were taken and analyzed by using XLSTAT computer software in one-way ANOVA, and 5% Dancun's multiple range was used for comparison. The results showed that the highest (86.66%) rooting was achieved in the cuttings treated with 6 g/L licorice extract and they were significantly different with control cuttings (53.33%), but they were not significantly different with 3000 ppm IBA (66.66%). Cinnamon 3g/L and fenugreek 3g/L extracts gave the lowest (6.66% and 33.33%, respectively) rooting and other studied parameters. The cuttings dipped in 3000 ppm IBA gave the highest (18.91) root number and the highest (66.66%) survival cuttings after transplanting. The longest root (15.54 cm) was found in cuttings were treated with 6 g/L moringa extract. The longest (5.83 cm) shoot was observed in treated cuttings with 3 g/L willow extract. The highest chlorophyll a and b (10.08 and 4.62 mg/L, respectively) were observed in cuttings treated with 6 g/L willow extract. Moreover, 3000 ppm IBA gave the highest (20.23%) total carbohydrate and (1.77 mg/g) IAA content along with 6 g/L licorice, moringa and fenugreek extracts, after 30 days from planting of the cuttings. Licorice root extract at 6 g/L fairly improved the measurements similar to 3000 ppm IBA throughout the study.

**Introduction**

Bottlebrush (*Callistemon viminalis*) is an ornamental evergreen shrub from Myrtaceae family which is attractive by its falling branches and red flowers in cylinder spikes [31]. Bottlebrush is propagated by seed or stem cuttings, however [22] preferred stem cuttings for propagation bottlebrush over seeds because the seedlings from seeds are worthless as ornamentals. Stem cuttings of bottlebrush are relatively difficult-to-root, and root induction in the cuttings requires hormonal and environmental treatments [38]. Despite the significant role of synthetic auxins in inducing rooting in callistemon cuttings, but natural rooting promoters instead of auxins are used since synthetic auxins are expensive, not easily available and hazardous for human [11]. In this regard, plant extracts are used to stimulate growth in plants and to induce formation of adventitious roots in propagules including cuttings. Natural occurring compounds in many plant extracts have been noticed by a number of researchers which induce rooting in cuttings. For example, more than 100 constituents are found in licorice extract, main of them are glycyrrhizin, phenolic compounds, mevalonic acid, polysaccharide, lignin, asparagine, vitamins and pantothenic acid which in whole induce the plant growth [8;18]. Also, willow extract is another alternative used to stimulate adventitious root formation in cuttings [7]. Willow is rich in salicylic acid which its role in ARF was found in mung bean (*Phaseolus radiatus* L) hypocotyl cuttings [37]. Moreover, higher root characteristics in cuttings of *Parkia biglobosa* were observed when treated with moringa leaf extract [16].

On the other hand, concentration of rooting promoter substances directly influences the rooting ability of cuttings. Variation in rooting response of cuttings occurred with change in applied auxins concentration [19]. For plant extracts, [25] mentioned the concentration of kelp extract as a key factor for its effectiveness. In view of these, this research is selected to investigate effect of some plant extracts in different concentrations on bottlebrush (*Callistemon viminalis*) hardwood cuttings.

## Materials and Methods

The study was carried out in a greenhouse at the Collage of Agricultural Engineering Sciences, University of Sulaimani, Kurdistan Region-Iraq during the year (2021) to study the effects of some plant extracts on hardwood cuttings of bottlebrush (*Callistemon viminalis*) in comparison with control and IBA treatments.

## Preparation of plant extracts and IBA

The extraction was started by powdering the plant materials in a blender, from moringa leaf, licorice root, willow leaf and new shoots, fenugreek seed and cinnamon bark. The powder of each of the plant was weighed two times, 3 and 6 grams, and placed in conical flask, and then distilled water was added to the required volume (1 liter) and sealed. The conical flasks were placed in a water bath at temperature (39 ºC) for 2 hours, after that left in dark for 24 hours. In the next day, the mixtures of the plants were filtered through a thin cloth, and the filtered liquids were used for treating the cuttings [4]. While, the IBA solution was prepared by dissolving it in 50% ethanol with a concentration of 3000 ppm.

## Treatment and planting of cuttings

The cuttings were collected on 11 March 2021 from basal part of one-year-old shoots with 20 cm long of 14-year-old bottlebrush trees. The cuttings, after preparation, were divided into lots with 18 cuttings, and each lot was separately treated with plant extract concentrations, 3 and 6 g/L, for 1 hour, and 3000 ppm IBA for 10s, but control cuttings was just dipped in distilled water for 1 hour. After treatment, the cuttings were planted in a mixture of sand and rice husk medium which was placed in polyethylene bags with a size of 15×35 cm. Six cuttings were planted in each bag, and the bags laid out in Complete Randomized Design (CRD) with three replications in a greenhouse. The average maximum and minimum temperature inside the greenhouse was between 13-29ºC

## Chemical measurements

Thirty days after planting of the cuttings, one cutting was taken from each replication of each treatment for quantification of total carbohydrate and Indole-3-acetic acid (IAA) content of the planted cuttings.

## Total carbohydrate measurement

The total carbohydrate was quantified according to Lane-Eynon method which described by [6].

## Extraction and analysis of indole-3-acetic acid

(3 gm) were weighed and was crushed in a mortar and pestle and extracted with (20 ml) methanol. The extract was cleared by centrifugation 5000 rpm /10 min. The resulting supernatant was transferred to a new tube. The pH of the plant extract was adjusted higher than 9 ml with 1M KOH to keep IAA. To increase the polarity of the sample before partitioning against ethyl acetate, one volume of pure water was added (2 ml) of ethyl acetate and (5 ml) of distilled water was added to the final extract The aqueous and organic phases were separated by centrifugation 5000 rpm / min, and the lower aqueous phase was transferred to a new tube [26].

HPLC Condition:

Mobile phase = methanol: 2% acetic acid (70:30)

Column = C18 – ODS (25 cm * 4.6 mm)

Detector = UV – 273 nm

Flow rate = 1.2 ml/ min

**Estimation chlorophyll content**

Leaf chlorophyll a and b of the cuttings were determined spectrophotometrically at 646.8 and 663.2 nm in 80% acetone as described by [32].

**Data Collection and Statistical Analysis**

Cuttings from the experiment were uprooted 4 months after planting for evaluating the data. The data were rooting percentage, number of roots, the length of the longest root, shoot length and shoot diameter. After taking these data, the cuttings were transplanted in polyethylene bags filled up with mixture of soil and compost and place in a greenhouse in which the temperature was between (40-18 °C) for calculating survival percentage after planting, and 30 days later it was calculated. The collected data were analyzed by using XLSTAT computer software in on-way ANOVA-CRD, and 5% Dancun's multiple range used for comparison of means.

**Results and discussion**

The criteria in this study were the results of control cuttings and cuttings dipped in 3000 ppm IBA which the results of treated cuttings with the plant extracts were compared.

**Rooting percentage**

The results in table (1) showed that different rooting percentages were achieved in the hardwood cuttings of *callistemon viminalis* when they were treated with different concentrations (3 and 6 g/L) of the plant extracts. Rooting percentage was significantly different in the cuttings treated with 6 g/L licorice extract in comparison with control cuttings; the same treated cuttings gave no different rooting percentage compared with the cuttings were dipped in 3000 ppm IBA. The highest (86.66%) rooting was achieved in the cuttings treated with 6 g/L licorice extract, control cuttings gave (53.33%) rooting and the cuttings dipped in 3000 ppm IBA gave (66.66%) rooting, but the lowest (6.66% and 33.33) rooting observed in the cuttings treated with 3 g/L cinnamon and fenugreek extracts, respectively. Improving rooting in the cuttings were treated with 6 g/L licorice extract may be due to that licorice contains many chemical compounds which may have positive effect on adventitious root formation (ARF) in cuttings. [17] referred that licorice extracts improved root quality in grape cuttings near to IBA. They explained this is as a result of containing combination of many components, minerals and growth regulators in licorice extract that contributed to boosting cuttings sprouting, rooting and growth. The noticeable components in licorice are flavonoids, triterpenoids, polysaccharides and phenolic compounds [30]. In this context, quercetin is a flavonoid exists in licorice extracts [33] which has role in ARF in cuttings [15]. [34] found that *Ilex paraguariensis* cuttings treated with aqueous solution of 30 μM quercetin gave three times more rooting percentage than control cuttings. The phenolic compounds also occur in licorice extract [36], and they may affect ARF in the plant parts as propagule. [14] found in stem slice cut from apple microshoots that phenolic compounds enhanced ARF because of they act as protective antioxidants. Additionally, licorice extract may have capacity to act as antifungal and protect cuttings from infection with fungi. It was observed that licorice extract decreased growth rate of deleterious fungi in media of mycelia of *Pleurotus ostreatus* [5]. Additionally, inhibitory effect of cinnamon extract on rooting of *Melaleuca viminalis* L. was reported by [20] at concentration of 4 mg 100 mL$^{-1}$ without combination with auxins.

Table 1: Effect of 3000 ppm IBA and different plant extract concentrations on rooting percentage, root number and root length of hardwood cuttings of *Callistemon vimnalis*.

| Treatments | Rooting % | Root No. | Root length (cm) |
| --- | --- | --- | --- |
| Control | 53.33 bcd | 6.16 b | 8.14 ab |
| IBA 3000 ppm | 66.66 abc | 18.91 a | 12.40 a |
| Licorice 3g/L | 60 a-d | 5.44 b | 10.43 ab |
| Licorice 6g/L | 86.66 a | 5.65 b | 12.31 a |
| Moringa 3g/L | 73.33 ab | 4.36 bc | 11.41 a |
| Moringa 6g/L | 53.33 bcd | 4.36 bc | 15.54 a |
| Willow 3g/L | 53.33 bcd | 4.5 bc | 12.19 a |
| Willow 6g/L | 73.33 ab | 4.04 bc | 8.45 ab |
| Fenugreek 3g/L | 33.33 de | 3.66 bc | 7.36 ab |
| Fenugreek 6g/L | 46.66 bcd | 5 b | 12.58 a |
| Cinnamon 3g/L | 6.66 e | 0.66 c | 0.40 b |
| Cinnamon 6g/L | 40 cd | 3.8 bc | 9.68 ab |

* The values in each column with the same letter do not differ significantly ($P \leq 0.05$) according to Duncan's Multiple Range Test.

**Root number and length**

Table (1) also demonstrated effect of the plant extracts on root number, according to the results the plant extracts did not increase root number over control cuttings and the cuttings dipped in 3000 ppm IBA. In opposite, the cuttings dipped in 3000 ppm IBA significantly increased root number. The cuttings dipped in 3000 ppm IBA gave the highest (18.91) root number, while the lowest (0.66) root number observed in the cuttings treated with 3 g/L cinnamon extract. Increasing root number in the cuttings dipped in 3000 ppm IBA could be interpreted as that IBA may cause enhancing the enzyme content of cuttings and increasing cell division, hence it increases root number. [9] referred that high level of endogenous enzymes intensified root number in *Dendrobium nobile* microshoots on account of auxin application. [27] reported that cell division gave rise to lateral root formation as a consequence of treatment with auxin. Moreover, IBA may produce more roots because it makes great amount of endogenous auxin content, stabilizes endogenous auxin against catabolism and inactivates growth inhibitors through their conjugate formation [29]. Auxins induce bud sprout in cuttings as well [21], and [31] reported that emergence new shoots and leaves before root formation on the cuttings raise photosynthesis and carbohydrate assimilation which in turn increase additional new roots.

On the other hand, root length in the present study was sharply decreased in the cuttings were treated with 3 g/L cinnamon extract in comparison with control cuttings, those dipped in 3000 ppm IBA and the cuttings treated with the rest of other plant extracts (Table 1). The longest root (15.54 cm) was found in the cuttings were treated with 6 g/L moringa extract, but the shortest root (0.40) was found in the cuttings treated with 3 g/L cinnamon extract. Elongation of roots in the cuttings were treated with 6 g/L moringa extract may belong to that moringa extract contains cytokinin zeatin [12], and [16] found in semi-hardwood stem cutting of *Parkia biglobosa* that moringa leaf extract gave higher total root length and length of the longest root due to cytokinin zeatin presence which promotes cell division.

**Shoot length and diameter**

The results of table (2) revealed that shoot length was not statistically different in control cuttings, those dipped in 3000 ppm IBA and the cuttings were treated with the five plant extracts, excluding 3 g/L cinnamon extract. The cuttings treated with 3 g/L cinnamon extract reduced shoot length. The longest (5.83 cm) shoot was observed in treated cuttings with 3 g/L willow extract, and the shortest (0.53 cm) was observed in treated cuttings with 3 g/L cinnamon extract. Shoot diameter was not different significantly in the cuttings treated with the plant extracts, 3000 ppm IBA and control cuttings (Table 2). Elongation new formed shoots in the cuttings treated with 3 g/L willow in the present study may be due to willow rich in salicylic acid [13], and salicylic acid may ameliorate the metabolism of energy storage compounds such as carbohydrates and proteins, and releases more energy for better growth. It was revealed that application of salicylic acid increased soluble proteins, free amino acids and sugars content of microcuttings of *Vaccinium corymbosum* [28]. The metabolites are source of energy for cell division and growth in cuttings [23]. While, the shortest shoot due to application of the cinnamon extract at 3 g/L may be related to that this concentration gave poor root traits, and a vigorous root system is required to vigorous shoot growth.

Table 2: Effect of 3000 ppm IBA and different plant extract concentrations on shoot length shoot diameter and survival percentage after transplanting of hardwood cuttings of *Callistemon vimnalis*.

| Treatments | Shoot length (cm) | Shoot diameter (mm) | Survival percentage after transplanting % |
|---|---|---|---|
| Control | 2.91 abc | 1.40 a | 26.66 d |
| IBA 3000 ppm | 4.17 ab | 1.65 a | 66.66 a |
| Licorice 3g/L | 3.30 abc | 1.28 a | 53.33 abc |
| Licorice 6g/L | 3.59 abc | 1.10 a | 60 ab |
| Moringa 3g/L | 3.96 ab | 1.35 a | 53.33 abc |
| Moringa 6g/L | 4.72 ab | 1.45 a | 26.66 d |
| Willow 3g/L | 5.83 a | 1.86 a | 40 bcd |
| Willow 6g/L | 3.67 abc | 1.33 a | 33.33 cd |
| Fenugreek 3g/L | 1.66 bc | 0.82 a | 26.66 d |
| Fenugreek 6g/L | 4.85 ab | 1.45 a | 33.33 cd |
| Cinnamon 3g/L | 0.53 c | 0.92 a | 0 e |
| Cinnamon 6g/L | 3.08 abc | 1.05 a | 20 de |

* The values in each column with the same letter do not differ significantly ($P \leq 0.05$) according to Duncan's Multiple Range Test.

**Survival percentage after transplanting**

The data of survival percentage after transplanting (Table 2) explained that the cuttings were treated with both concentrations of licorice extracts, 3 and 6 g/L, and cuttings treated with 3 g/L moringa extracts gave better percentage number of the cuttings survived after transplanting than control cuttings and other cuttings treated with the rest of the plant extracts, but they were not different with cuttings dipped in 3000 ppm IBA. Cuttings were dipped in 3000 ppm IBA gave the highest (66.66%) survival cuttings, whereas 6 and 3 g/L of licorice extracts and 3 g/L moringa extract gave (60%, 53.33% and 53.33%, respectively) survival cuttings after transplanting. No cuttings were survived after transplanting as they treated with 3 g/L cinnamon extract. Cuttings were dipped in 3000 ppm IBA gave the best root number, and these might give rise to the highest survived cuttings. High number of roots gives the transplanted cuttings more water and nutrients, thus the transplanted cuttings have more opportunities to survive in a high rate. Similarly, [10] argued that a better growth of rooted cuttings after transplanting belonged to a large number of roots in the cuttings.

**Estimation chlorophyll content**

The data in table (3) showed that the plant extracts with the exception of 3 g/L cinnamon extract produced no noticeable differences in chlorophylls ratio in hardwood cuttings of bottlebrush when used as rooting promoter. However, 3 g/L cinnamon extract diminished chlorophylls content in the cuttings compared with control cuttings and the cuttings dipped in 3000 ppm IBA. The highest (10.08 mg/L) chlorophyll a was obtained from the cuttings soaked in 6 g/L willow extract, control cuttings gave (9.9 mg/L), and the cuttings treated with 6 g/L licorice extract gave (9.43 mg/L). Also, the same cuttings gave the best chlorophyll b content, and it was superior (4.62 mg/L) in the cuttings soaked in 6 g/L willow extract. In contrast, 3 g/L cinnamon gave the lowest amount of both the chlorophylls a and b (2.06 and 0.93 mg/L, respectively). As mentioned above, salicylic acid exists in willow extract; it was found that salicylic acid has role in raising chlorophyll content in leaves of cuttings. [3] found in cuttings of *Amygdalus* L. that acetylsalicylic acid (ASA) which is a derivative of salicylic acid improved chlorophyll content in association with IAA.

Table 3: Effect of 3000 ppm IBA and different plant extract concentrations on chlorophyll a and b of hardwood cuttings of *Callistemon vimnalis*.

| Treatments | Chlorophyll a mg/L | Chlorophyll b mg/L |
|---|---|---|
| Control | 9.90 a | 4.48 a |
| IBA 3000 ppm | 8.20 ab | 3.55 ab |
| Licorice 3g/L | 6.88 ab | 2.91 ab |
| Licorice 6g/L | 9.43 a | 3.96 a |
| Moringa 3g/L | 5.54 ab | 2.30 ab |
| Moringa 6g/L | 7.83 ab | 3.46 ab |
| Willow 3g/L | 7.81 ab | 3.66 ab |
| Willow 6g/L | 10.08 a | 4.62 a |
| Fenugreek 3g/L | 6.61 ab | 2.86 ab |
| Fenugreek 6g/L | 7.27 ab | 2.74 ab |
| Cinnamon 3g/L | 2.06 b | 0.93 b |
| Cinnamon 6g/L | 6.87 ab | 3.17 ab |

\* The values in each column with the same letter do not differ significantly (*P*≤0.05) according to Duncan's Multiple Range Test.

**Total carbohydrate**

The data in table (4) exhibited total carbohydrate in callistemon hardwood cuttings that were taken 30 days after planting. According to the data, total carbohydrate was not increased in the cuttings treated with the plant extracts compared with control cuttings and the cuttings dipped in 3000 ppm IBA. Dipping cuttings in 3000 ppm IBA increased total carbohydrate significantly compared with control cuttings and the cuttings treated with 3 g/L extracts of moringa, willow and fenugreek.

The cuttings dipped in 3000 ppm IBA gave the highest (20.23%) total carbohydrate, to a lesser extent 6 g/L of licorice extract showed (18.45%) total carbohydrate. Control cuttings gave the lowest (13.05%) total carbohydrate. Also, the percentage of total carbohydrate in the cuttings treated with 3 g/L moringa, willow and cinnamon extracts were low and their results were near to the result of control cuttings. Similarly, cuttings of *Myrtus communis* demonstrated highest total carbohydrate when they were dipped in 4000 ppm IBA [1]. Total carbohydrate may play role as a source and storage to provide metabolic energy in the following stages of ARF formation, and growth regulators from root promotor groups are likely helpful to metabolize and mobilize carbohydrate content to the place of root formation in cuttings. [24] indicated that auxins, which are rooting promoter in cuttings, are of vital importance to mobilization carbohydrates from leaves, upper stem parts and concomitantly increase sugar content at rooting zone through hydrolysis of starch.

Table 4: Effect of 3000 ppm IBA and different plant extract concentrations on total carbohydrate and IAA content of hardwood cuttings of *Callistemon vimnalis* after 30 days from planting date.

| Treatments | Total carbohydrate % | IAA content mg/g |
|---|---|---|
| Control | 13.05 b | 0.19 e |
| IBA 3000 ppm | 20.23 a | 1.77 a |
| Licorice 3g/L | 15.98 ab | 1.17 abc |
| Licorice 6g/L | 18.45 ab | 1.32 ab |
| Moringa 3g/L | 13.89 b | 0.89 de |
| Moringa 6g/L | 15.79 ab | 1.51 ab |
| Willow 3g/L | 13.78 b | 0.5 de |
| Willow 6g/L | 14.56 ab | 1.21 abc |
| Fenugreek 3g/L | 13.98 b | 0.89 bcd |
| Fenugreek 6g/L | 16.41 ab | 1.41 ab |
| Cinnamon 3g/L | 14.65 ab | 0.65 cde |
| Cinnamon 6g/L | 16.59 ab | 1.11 bcd |

* The values in each column with the same letter do not differ significantly ($P\leq0.05$) according to Duncan's Multiple Range Test.

**Indole-3-acetic acid (IAA) content**

Table (4) also depicted the IAA content in the cuttings 30 days after planting of the cuttings. As shown, the cuttings treated with the plant extracts raised IAA content, and they were at variance with control cuttings except for the cuttings treated with 3 g/L extracts of willow and cinnamon. All plant extract treatments were unable to produce significant IAA content when compared with the 3000 ppm IBA treated cuttings. The highest (1.77 mg/g) IAA detected in cuttings dipped in 3000 ppm IBA. IAA content reached (1.51 mg/g) in the cuttings treated with 6 g/L moringa extract. Concomitantly, extracts of fenugreek, licorice, and willow at 6 g/L gave (1.41, 1.32 and 1.21 mg/g, respectively) IAA content, they were statistically similar to 3000 ppm IBA. Control cuttings gave the lowest (0.19 mg/g) IAA content, 3 g/L cinnamon extract (0.65 mg/g) as well. Researcher found that exogenous application of IBA to the cuttings induces increasing IAA content of the cuttings via its conversion to IAA or enhancing biosynthesis of IAA. [35] found that applied IBA converted to IAA in stem slices of apple shoots, also they referred that IBA may change action or endogenous synthesis of IAA. Besides, [2] concluded that exogenous spray of moringa extracts increased plant hormones including auxin in rocket plants. Additionally, licorice extracts produced the best IAA content in grape cuttings along with IBA [17].

**Interdependence of the measured characteristics**

According to the results demonstrated in figure (1), rooting percentage positively correlated with root length (r = 0.67, p-value = 0.017), shoot length (r = 0.59, p-value = 0.044), survival percentage after transplanting (r = 0.84, p-value = 0.001), and chlorophyll a and b (r =

Euphrates Journal of Agriculture Science-13 (3): 89-98, (2021)   Mustafa et al.

072 and 0.66, p-value = 0.008 and 0.019, respectively). Root number positively associated with survival percentage after transplanting (r = 0.67, p-value = 0.017), and root length strongly correlated with shoot length (r = 0.87, p-value = 0.0002), shoot diameter (r = 0.61, p-value = 0.034) and survival percentage after transplanting (r = 0.64, p-value = 0.026). The results of root trait correlations confirmed the previous statements that the best root traits lead to better shoot growth and the highest survival percentage after transplanting of the cuttings. This is because of the highest root number and root length mean absorbing more water and nutrients.

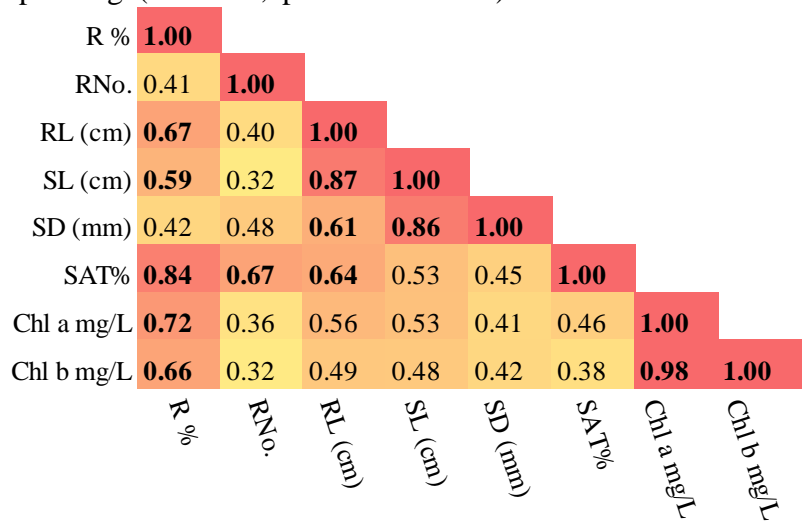

Figure 1: Pearson correlation analysis of the studied parameters. R%: rooting percentage, RNo.: root number, RL: root length, SL: shoot length, SD: shoot diameter, SAT%: survival percentage after transplanting, Chl a: chlorophyll a, Chl b: chlorophyll b.

## Conclusion

The results of application 6 g/L licorice root extract as rooting promoter to the hardwood cuttings of bottlebrush were similar to the results of application 3000 ppm IBA in the most studied parameters. So, it could be possibly concluded that 6 g/L licorice root extract will be efficiently used for promoting root formation in the cuttings. This was more evident in the two key studied characteristics rooting percentage and survival percentage after transplanting.